\documentclass[%
 reprint,
superscriptaddress,
 amsmath,amssymb,
 aps
]{revtex4-2}

\usepackage{graphicx}
\usepackage{dcolumn}
\usepackage{bm}
\usepackage{hyperref}
\usepackage[mathlines]{lineno}
\usepackage{color}

\usepackage[normalem]{ulem}

\newcommand{\Eqref}[1] {Eq.~\ref{#1}}
\newcommand{\Figref}[1] {Fig.~\ref{#1}}

\newcommand{\appref}[1] {Appendix~\ref{#1}}
\newcommand{\secref}[1] {Section~\ref{#1}}

\newcommand{\n}[0]{\nonumber\\}

\newcommand{\kt}  [0] {k_\mathrm{B} T}
\newcommand{\omn} [0] {\omega_0}

\newcommand{\cte}[0] {\mathrm{cte}}

\newcommand{\npl}[0] {\langle n \rangle}

\newcommand{\lk}	[0] {\mathrm{Lk}}
\newcommand{\dlk}	[0] {\Delta \lk}
\newcommand{\lkn}	[0] {\lk_0}

\newcommand{\sig} [0] {\sigma}
\newcommand{\sigs} [0] {\sigma_s}
\newcommand{\sigp} [0] {\sigma_p}

\newcommand{\ffe}[0]{{\cal F}}
\newcommand{\sfe}[0]{{\cal S}}
\newcommand{\pfe}[0]{{\cal P}}

\newcommand{\tffe}[0]{\widetilde{\ffe}}
\newcommand{\tsfe}[0]{\widetilde{\sfe}}
\newcommand{\tpfe}[0]{\widetilde{\pfe}}

\renewcommand{\vec} [1] {\mathbf{#1}}

\begin{document}


\title{Statistical Mechanics of Multiplectoneme Phases in DNA}


\author{Midas Segers}
\thanks{These two authors contributed equally}
\affiliation{Soft Matter and Biophysics, KU Leuven, Celestijnenlaan 200D, 3001 Leuven, Belgium}
\author{Enrico Skoruppa}
\thanks{These two authors contributed equally}
\affiliation{
    Cluster of Excellence Physics of Life, TU Dresden, 01062 Dresden, Germany
}
\author{Helmut Schiessel}
\affiliation{
    Cluster of Excellence Physics of Life, TU Dresden, 01062 Dresden, Germany
}
\affiliation{
    Institut f{\"u}r Theoretische Physik, TU Dresden, 01062 Dresden, Germany
}
\author{Enrico Carlon}%
\affiliation{Soft Matter and Biophysics, KU Leuven, Celestijnenlaan 200D, 3001 Leuven, Belgium}


\date{\today}


\begin{abstract}
A stretched DNA molecule which is also under- or overwound, undergoes a
buckling transition forming intertwined looped domains called plectonemes.
Here we develop a simple theory that extends the two-phase model of
stretched supercoiled DNA, allowing for the coexistence of multiple
plectonemic domains by including positional and length distribution
entropies.  Such a multiplectoneme phase is favored in long DNA
molecules in which the gain of positional entropy compensates for
the cost of nucleating a plectoneme along a stretched DNA segment.
Despite its simplicity, the developed theory is shown to be in excellent
agreement with Monte Carlo simulations of the twistable wormlike chain
model. The theory predicts more plectonemes than experimentally observed,
which we attribute to the limited resolution of experimental data.
Since plectonemes are detected through fluorescence signals, those
shorter than the observable threshold are likely missed.
\end{abstract}

\maketitle


\section{Introduction}
\label{sec:intro}

\begin{figure}[b]
\centering\includegraphics[width=0.48\textwidth]{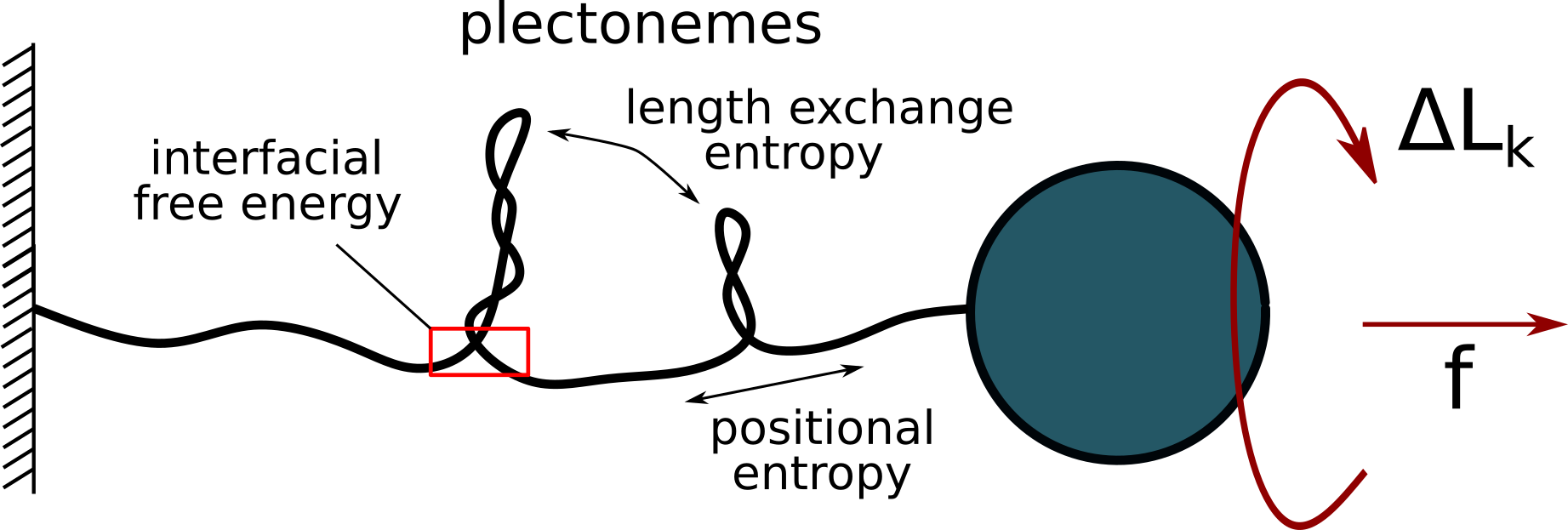}
\caption{Setup of a typical DNA Magnetic Tweezers experiment. The DNA
molecule is tethered at one end to a solid surface and at the other to
a paramagnetic bead. A magnetic field is applied which exerts a linear
force $f$ and rotates the bead to a $\Delta Lk$ number of turns (counted
with respect to the torsionally relaxed state). When $\Delta Lk$ exceeds
a threshold value the molecule buckles and plectonemic supercoils appear.
We develop a theory of multiplectonemes that incorporates a free energy
penalty for plectoneme nucleation, along with two entropic terms: one
for positional entropy and another for length-exchange entropy.}
\label{fig:intro} 
\end{figure}

In the cellular environment, DNA typically exists in a
torsionally underwound state, a conformation finely regulated
by the concerted activity of enzymes such as topoisomerases and
polymerases~\cite{Koster2010}. Over- and underwound DNA—also
called supercoiled DNA—exhibits a distinct response to the imposed
torsional strain that culminates in the assumption of superhelically
coiled configurations called plectonemes~\cite{bole90,mark95b}
(from the Greek {\sl{pl\'eko}}, braid and {\sl{n\'\i ma}},
filament; see \Figref{fig:intro}).  The transition into the
plectonemic state is commonly referred to as DNA buckling.
Plectonemic supercoiling is known to be involved in a multitude
of biological processes, notably in the regulation of gene
expression~\cite{Dorman2016,MartisB2019,Johnstone2022,Patel2023},
the maintenance of chromatin
architecture~\cite{Deng2005,Gilbert2014,corl16,blos17,schi21,Guo2021},
and retroviral integration~\cite{vand19}. One mechanism by which
supercoiling promotes biological function is by bringing distal sites
in close proximity to favor, for example, the binding of DNA-bridging
proteins~\cite{yan18a,yan18b,yan21,vand19,vand22} and the interaction
between promoter-enhancer pairs~\cite{liu2001,bene2014}. The
broad phenomenology of DNA supercoiling has been
studied extensively due to its relevance in biology~\cite{stri96,
fort08,wada09,neuk11,vanl12,fath15,lepa15,fort19,walt21,fosa21,pyne21,vand22,skor22,wats22,juni23},
yet many aspects remain elusive. Whether or not supercoiling is a
plausible mechanism leveraged in the establishment of DNA bridges
critically depends on the relation between the distance of the binding
motives and the typical size of plectonemes. If the distance between these
sites far exceeds the characteristic size of a plectoneme, supercoiling
is unlikely to contribute to the establishment of proximity between
the relevant segments.  Conversely, binding between motives within
the range of characteristic plectoneme lengths is likely enhanced by
supercoiling. In this work, we explore to which degree supercoiling
is segregated into multiple plectonemes and how this distribution is
regulated by external forces and torques.

We study this phenomenology in the context of single-molecule
magnetic tweezers (MT)~\cite{smit92,mosc09,lipf10}, which have
emerged as prominent tools for the experimental exploration of DNA
supercoiling. In MT experiments, a single DNA molecule is tethered between
a superparamagnetic bead and a flowcell surface. Exposure of the bead
to an appropriately calibrated magnetic field allows for the induction
of linear stretching forces (see \Figref{fig:intro}).  Additionally,
rotating the field-inducing magnet enables the torsional state of the
molecule to be controlled.  This torsional state corresponds to the mutual
wrapping of the two individual strands of a double-stranded DNA molecule.
In the torsionally relaxed state, these strands wrap around each other
approximately once every $10.5$ base pairs, resulting in a relaxed state
linking number of approximately $\lkn \approx N/10.5$ for a molecule
consisting of $N$ base pairs.  Rotation of the bead induces an excess
linking strain $\dlk = \lk -\lkn \neq 0$.

The classical readout of these experiments is limited to the
tether extension (i.e., the bead-surface distance), which
may be observed in terms of the imposed number of turns on the
bead. Plectoneme formation becomes indirectly visible due to the
significant and progressive reduction in tether extension with
increasing linking strain. This behavior—characterized by the mean
extension~\cite{mark95b,moro98,mark07,mosc09,mark15,lipf10,neuk11}
and extension fluctuations~\cite{vand22,skor22}—are well understood.
However, this readout does not yield access to the morphology and phase
characteristics of the underlying molecule. Several studies have attempted
to shed light on these characteristics by combining MT (or similar)
setups with fluorescent microscopy, which enables the visualization of
the molecule~\cite{vanl12,shep24}. Van Loenhout et al.~\cite{vanl12}
demonstrated the propensity of DNA to nucleate into more than one
plectoneme.  In the present work, we revisit this observation by
leveraging a combination of simulation and analytical approaches.

In theoretical treatments of DNA supercoiling, introducing the
supercoiling density $\sigma = \dlk /\lkn$ proves convenient as it is
independent of the molecule's curvilinear length. Due to its chiral
nature, DNA exhibits a different response to overwinding ($\sigma >
0$) versus underwinding ($\sigma < 0$). Notably, underwinding, coupled
with adequate stretching forces, leads to a structural transition
known as torsionally induced melting \cite{shei11}. However, this
transition predominantly occurs for stretching forces exceeding
$1$~pN~\cite{stri96}. Here, we restrict our analysis to the linear
elastic regime, ignoring higher-order transitions, thus rendering our
approach fully symmetric regarding over- and underwinding.

DNA buckling is commonly described as a (pseudo) first-order phase
transition \cite{mark07}.  While gradually increasing $\sigma$ starting
from the torsionally relaxed state ($\sigma=0$), the tethered molecule
initially remains in an elongated state until a threshold value $\sig =
\sigs$ is reached. Beyond this (buckling) point plectonemes nucleate along
the stretched phase, as depicted in  Fig.~\ref{fig:intro}. This transition
comes at the cost of a considerable reduction in tether extension but
enables the absorption of more linking strain due to the crossings
within the plectonemes, i.e., the supercoiling density of plectonemes
exceeds that of the stretched phase ($\sigma_p > \sigma_s$). This
phenomenon resembles liquid-vapor coexistence, where the two phases are
maintained at different particle densities ($n_{liq.} > n_{vap.}$).

In this work, we employ efficient Monte Carlo simulations (MC) of a
discrete twistable wormlike chain (TWLC) to infer physical properties
of plectonemes in stretched and torsionally constrained DNA.  Moreover,
we develop a simple statistical mechanical approach that describes a
DNA molecule as a chain consisting of stretched and plectonemic phase
segments.  The model features a free energy penalty associated with
the interface between the two phases that tends to suppress plectoneme
nucleation.  Nevertheless, in long DNA molecules, the nucleation of
multiple plectonemes becomes entropically favorable. We account for two
types of entropic contributions: positional entropy, which reflects the
number of possible positions for plectonemes along the stretched segments,
and length-exchange entropy, which describes the partitioning of the total
plectonemic length among different plectonemes (see Fig.~\ref{fig:intro}),
similar to \cite{eman13}.  Throughout this work, we will refer to this
model as the multiplectoneme model (MP).  Our approach is simpler than
existing more sophisticated theories of the multiplectonemic phase
considered in the past \cite{neuk11,eman13}. Despite its simplicity,
the model displays excellent agreement with MC-generated data regarding
the average number of plectonemes, their lengths and distribution, as
well as other quantities such as the torque versus supercoiling density.
Moreover, we demonstrate the model to reproduce additional features
observed in a recent MT study~\cite{gao21}.

This paper is organized as follows: \secref{sec:theory} reviews the
two-phase model of stretched DNA buckling and introduces the theoretical
framework of the multiplectoneme model.  \secref{sec:MC} introduces
the Monte Carlo simulations and the methodology of inference for the
relevant observables. The simulation readout is then compared to the model
predictions for various quantities. Contextualization and comparison
with experimental data are provided in \secref{sec:expt}. Finally,
\secref{sec:conclusion} concludes the paper by highlighting the relevance
of our findings.


\section{Modeling the multiplectoneme state}
\label{sec:theory}

In the two-phase model of stretched DNA buckling, a DNA molecule of
length $L$, subject to a stretching force $f$ and at fixed supercoil
density $\sigma$ is described as consisting of two distinct phases
\cite{mark95b,mark07}: the stretched phase and plectonemic phases,
with corresponding free energies per unit lengths $\sfe(\phi)$ and
$\pfe(\psi)$, respectively, where $\phi$ and $\psi$ are the supercoil
densities in the two phases.

For the introduction of the theoretical description, we will restrict the
discussion to simple quadratic free energies, that allow for closed-form
expressions for many relevant quantities and therefore serve for the
purpose of illustration.  When comparing to Monte Carlo-sampled and
experimental data, we will incorporate higher-order corrections that
have been shown in previous work to provide superior agreement with MT
and MC data ~\cite{skor22} (see Appendix~\ref{app:quartic}). In this
quadratic description, the free energies of the two phases are given
by~\cite{mark07}
\begin{eqnarray}
    \label{eq:defS}
    {\cal S}(\phi) 
    &=& 
    -g(f) + a(f) \phi^2,
    \\
    \label{eq:defP}
    {\cal P}(\psi) 
    &=& 
    b \psi^2.
\end{eqnarray}
The free energy of the stretched phase, \Eqref{eq:defS}, can be
derived from the twistable wormlike chain (TWLC) as a high-force
expansion~\cite{moro97,moro98}. The theory yields the coefficients
\begin{eqnarray}
    \label{eq:gf_TWLC} 
    g(f) 
    &=& 
    f \left(1 - \sqrt{\frac{k_BT}{Af}} + \ldots \right),
    \\
    \label{eq:af_TWLC}
    a(f) 
    &=& 
    \frac{C}{2} \left(1-\frac{C}{4A}\sqrt{\frac{k_BT}{Af}} + \ldots\right)
    k_BT \omega_0^2,
\end{eqnarray}
where $A$ is the bending stiffness, $C$ is the twist stiffness and
$\omega_0$ is the intrinsic twist of the double helix. Throughout this
work we will use the values $A=40$~nm, $C=100$~nm, which were shown to
yield excellent agreement with MT data~\cite{vand22} and $\omega_0 =
1.75$ nm$^{-1}$.  The high force expansions \eqref{eq:gf_TWLC}, and
\eqref{eq:af_TWLC} are valid in the regime $k_B T/Af \ll 1$, which
corresponds roughly to $f > 0.5$~pN.  The plectonemic free energy
\eqref{eq:defP} is phenomenological and may be viewed as a lowest-order
expansion of the underlying free energy. The parameter $b$ is usually
expressed as
\begin{equation}
    b=\frac 1 2 P k_B T \omega_0^2,
    \label{eq:plec_quad_fe}
\end{equation}
where $P$ is referred to as the effective torsional stiffness of the
plectonemic phase \cite{gao21}, which by analogy to $A$ and $C$ is
expressed in units of length.  Assuming a fraction $\nu$ of the molecular
length to be contained in the stretched phase—and consequently a
fraction $(1-\nu)$ in the plectonemic phase—the total free energy can
be written as~\cite{mark07}
\begin{equation}
    \label{eq:maxwellconstr}
    \ffe(\phi,\psi,\nu) = \nu \sfe(\phi) + (1-\nu) \pfe(\psi).
\end{equation}
Minimization of $\ffe$ under the constraint of fixed total linking
number - or equivalently total supercoiling density $\sigma = \nu \phi +
(1-\nu) \psi$ - leads to a double tangent construction~\cite{mark07}.

In terms of the parameters of the free energies, one finds the 
phase-boundaries~\cite{skor22}
\begin{eqnarray}
    \sigma_s = \sqrt{\frac{b g}{a(a-b)}} ,
    &\qquad&
    \sigma_p = \sqrt{\frac{a g}{b(a-b)}},
    \label{eq:sigsp_quad}
\end{eqnarray}
which mark the limits of the regime of coexisting plectoneme and stretched
phases: $\sigs < |\sigma| < \sigp$.  At supercoiling densities smaller
than $\sigs$ ($|\sigma| < \sigs$), the molecule is in a pure stretched
phase, referred to as the pre-buckling regime. Conversely, supercoiling
densities exceeding $\sigp$ ($|\sigma| > \sigp$) mark a pure plectonemic
phase. Partial or fully plectonemic states are commonly referred to as
the post-buckling regime. This simple theory describes buckling as a
thermodynamic first-order transition.  Note that this description does
not account for the distribution of the two phases (corresponding to
more than one plectoneme) along the molecule.

\subsection{The torque ensemble}

\begin{figure}[t]
    \centering
    \includegraphics[width=7.74cm]{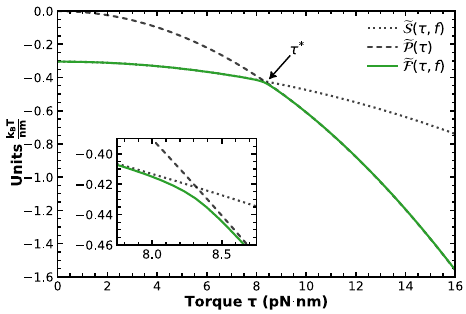}
    \caption{Free energy densities in the torque ensemble.  Dashed and
    dotted lines: free energies of the stretched \eqref{eq:Stilde}
    and plectonemic \eqref{eq:Ptilde} phases, respectively.  Green
    solid line: Free energy per unit length of the MP model, from
    Eq.~\eqref{def:F_hr}, vs. torque.  The inset provides a close-up
    view of the free energies near the buckling torque, highlighting the
    smooth transition captured by the MP model.  The parameters used for
    the graphs are $f=0.5$ pN, $\mu=50$~pN$\cdot$nm, and $l_c=100$ nm.
    The corresponding buckling torque (from Eq.~\eqref{eq:tau*}) is
    $\tau^{*} = 8.32$~pN$\cdot$nm.  }
\label{fig:free_en_tau}
\end{figure}

Entropically, the nucleation of multiple plectonemes is favored over
the formation of a single plectonemic domain. However, the propensity
for local phase-switches is constrained by free energy penalties
arising from interfacial tension. Consequently, the average number of
plectonemes is determined by the interplay between these two factors.
These effects cannot be accounted for by the two-phase theory introduced
above, since it only considers the balance between the two phases and
not their distribution along the chain.  For the incorporation of the
occurrence of multiple plectonemes in the thermodynamic description,
it turns out advantageous to consider the fixed torque ensemble ($\tau
= \cte$), instead of the fixed linking number ensemble ($\sigma =
\cte$). Although this ensemble differs from the scenarios typically
considered experimentally, we will still be able to deduce key observables
relevant to the linking number ensemble. In the context of MT experiments,
torque is related to the free energy via a partial derivative with
respect to $\theta$, the rotation angle described by the bead, which is
proportional to the supercoiling density $\theta = L \omega_0 \sigma$
\begin{equation}
    \label{eq:tau-sigma_ens}
    \tau = \frac{1}{\omega_0} \,\, \frac{\partial \ffe}{\partial \sigma},
\end{equation}
where ${\cal F}$ is the free energy per unit of length.  Stretched and
plectonemic phase free energies in the torque ensemble are related to
the linking number analogous via Legendre transforms. Transformation of
Eqs.~\eqref{eq:defS} and \eqref{eq:defP} yields
\begin{align}
    \label{eq:Stilde}
    \tsfe(\tau) &= -g - \frac{\omega_0^2}{4a} \, \tau^2,
    \\
    \label{eq:Ptilde}
    \tpfe(\tau) &= - \frac{\omega_0^2}{4b} \, \tau^2.
\end{align}
In the torque ensemble, the buckling torque is obtained from the condition
$\tsfe(\tau^*) = \tpfe(\tau^*)$ which gives
\begin{equation}
    \label{eq:tau*}
    \tau^* = \frac{2}{\omega_0}\sqrt{\frac{ab g}{a-b}}.
\end{equation}
The condition of equal torque implies the total free energy to be given
by the minimum of the two individual contributions
\begin{equation}
    \label{eq:minF}
    \tffe(\tau) 
    = 
    \min_\tau 
    \left( 
        \tsfe(\tau), \tpfe(\tau)
    \right).
\end{equation}
This means that in this approximation the molecule remains fully in the
stretched state for $|\tau| \leq \tau^*$ ($\tffe(\tau) = \tsfe(\tau)$) and
then immediately transitions into the fully plectonemic state for $|\tau|
\geq \tau^*$ ($\tffe(\tau) = \tpfe(\tau)$).  The dotted and dashed
lines in Fig.~\ref{fig:free_en_tau} are plots of the stretched and
plectonemic phase free energies for the quadratic model. The intersection
between these two curves corresponds to the buckling torque $\tau^*$.

\subsection{The multiplectoneme model}

We use the two-phase description of DNA supercoiling to obtain the free
energy associated with the multiplectonemic phase.  We distinguish between
three separate contributions to the multiplectonemic phase free energy:
one enthalpic and two entopic components.  Assuming a multiplectonemic
phase of $n$ plectonemes with a cumulative length $L_p$, the enthalpic
contribution to the free energy is given by
\begin{eqnarray}
\label{eq:enthalpy}
F_{enth.}(L,L_p,n) &=& (L-L_p) \, \widetilde{\mathcal{S}} + 
L_p \, \widetilde{\mathcal{P}} + n \mu,
\end{eqnarray}
where $\mu > 0$ is the free energy cost of inserting a plectoneme within
a stretched phase domain. This may be viewed as an effective elastic term
for the necessary bending to transition into a plectoneme and the creation
of a plectoneme end-loop.  A large value of $\mu$ results in a propensity
towards the formation of relatively few, but larger plectonemic domains,
whereas small values of $\mu$ favor the formation of a relatively large
amount of short plectonemes.  The nucleation free energy, $\mu$, serves
as a free parameter in our theory and will be fitted to Monte Carlo data,
as discussed in detail in Sec.~\ref{sec:MC}.

\begin{figure}[t]
	\centering
	\includegraphics[width=8.6cm]{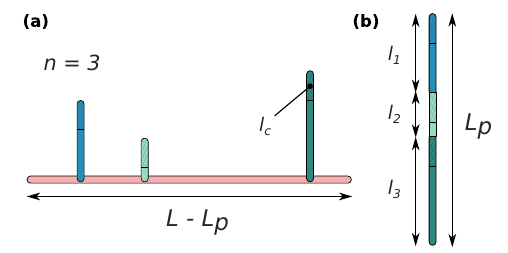}
	\caption{
	Schematic representation of the MP model for a multiplectonemic
	phase with $n=3$ plectonemes. (a) The configurational entropy
	\eqref{eq:entropy_config} quantifies the number of possible
	ways to distribute $n$ plectonemes along a stretched segment
	of length $L - L_p$.  (b) The number of ways the cumulative
	plectoneme length $L_p$ can be partitioned over the given
	number of plectonemes gives rise to the length-exchange entropy
	\eqref{eq:entropy_length}. The hatched regions indicate minimal
	size of the plectoneme $l_c$ in the MP model.}
	\label{fig:polydisperse_rod_schematic}
\end{figure}

The remaining entropic contributions to the free energy have been derived
previously in \cite{eman13}.  The first component is the positional
entropy. It counts the number of configurations in which the $n$
plectonemes can be inserted along the stretched phase (of length $L-L_p$)
\begin{equation}
        \Omega_n(L_p) =\frac{1}{n!}\left( \frac{L-L_p}{\Delta l} \right)^n,
	\label{eq:entropy_config}
\end{equation}
as illustrated in Fig.~\ref{fig:polydisperse_rod_schematic}(a). To make
$\Omega_n(L_p)$ dimensionless, we introduced a discretization length
$\Delta l$ which rescales all lengths. Technically, this introduces
configurations for which plectonemes occupy the same site. However,
for sufficiently small values of $\Delta l$, the statistical weight
of such configurations becomes vanishingly small.  Note, that this
expression is exact in the continuum limit. In practice, we choose a
discretization length of 1~nm. The second entropic component originates
from partitioning of the cumulative plectonemic length $L_p$ among $n$
different plectonemes. In order to discriminate small and non-physical
plectonemic domains we introduce a length cut-off $l_c$ which constitutes
the minimal size of a plectoneme. The partitioning of a segment of
total length $L_p$ into $n$ segments of minimal length $l_c$ consists in
splitting the interval $[0, L_p]$ into sub-intervals of lengths $l_1$,
$l_2$  \ldots $l_n$ with $\sum_k l_k = L_p$ and fulfilling the constraints
\begin{eqnarray}
 l_c       \leq \, &l_k& \, \leq L_p - (n-k)l_c - \sum_{m=1}^{k-1} l_m.
\end{eqnarray}
The number of ways in which this partitioning can be done is given by
\begin{equation}
  	\Lambda_n(L_p)=
  		\frac{1}{(n-1)!}
	\left( \frac{L_p-nl_c}{\Delta l} \right)^{n-1},
  	\label{eq:entropy_length}
\end{equation}
which corresponds to the partition function of a hard rod model.

Combining enthalpic \eqref{eq:enthalpy} and entropic
\eqref{eq:entropy_config}, \eqref{eq:entropy_length} terms and summing
up over all possible lengths of the plectonemic phase $L_p$ (at fixed
torque $\tau$) we obtain the total partition function for $n$ plectonemes
\begin{eqnarray}
Z(n) &=& \sum_{L_p} e^{-\beta F_{enth.}(L,L_p,n)} \Omega_n(L_p) \Lambda_n (L_p),
\label{def:Zn}
\end{eqnarray}
with $\beta = 1/k_BT$ and where the sum runs over values of $L_p$
which are multiples of the discretization length $\Delta l$ and with $n
l_c \leq L_p \leq L$.  Using the change of variable $x \equiv (L_p - n
l_c)/(L - n l_c)$ we can recast \eqref{def:Zn} in the following continuum
integral form
\begin{eqnarray}
Z(n) &=& \frac{\alpha_n}{n!(n-1)!} \int_0^1 e^{- \gamma_n x} (1-x)^n x^{n-1} dx,
\label{def:Zn-int}
\end{eqnarray}
where
\begin{eqnarray}
\alpha_n &\equiv & 
e^{-\beta F_{enth.}(L,n l_c,n)}
\left( \frac{L -n l_c}{\Delta l}\right)^{2n},
\end{eqnarray}
and 
\begin{eqnarray}
\gamma_n & \equiv & \beta (L-n l_c) 
\left( \widetilde{\cal P} - \widetilde{\cal S} \right).
\end{eqnarray}
Note that $\gamma_n$ is positive for $\tau<\tau^*$ and negative for $\tau>\tau^*$.

The total partition function of a DNA molecule of length $L$ 
is obtained by summing over the number of plectonemes $n$
\begin{eqnarray}
Z_\text{TOT}(L) &=&\sum_{n=0}^{n_\text{max}} Z(n)
\label{def:Ztot}
\end{eqnarray}
where the maximum number of plectonemes is set $n_\text{max} l_c= L$.  For
the range of lengths considered in this work—$L \approx 10~\mu\text{m}$
corresponding to $\sim 30$~kb—the sum in \eqref{def:Ztot} can be safely
truncated to $n \leq 20$ as the statistical weight of configurations
with larger numbers of plectonemes is negligible.  The free energy per
unit of length is then given by
\begin{eqnarray}
\widetilde{\cal F} = -\frac{k_BT}{L} \log Z_\text{TOT} (L).
\label{def:F_hr}
\end{eqnarray}
In practise, we calculate $Z_\text{TOT}$ by approximating
\eqref{def:Zn-int} via Laplace's method and evaluating the
summation in \eqref{def:Ztot} numerically.  The solid green line of
Fig.~\ref{fig:free_en_tau} is a plot of $\widetilde{\cal F}$ vs. the
torque $\tau$ for the MP model.  This free energy closely matches the
stretched and plectonemic free energy sufficiently far from the transition
region $\tau = \tau^*$, but it exhibits a smooth transition rather than
the sharp crossing predicted by the two-phase model.

To relate measurements carried out in the fixed linking 
number ensemble to predictions from our theoretical model calculated
within the fixed torque ensemble, we note that first moments in the 
two ensembles are equivalent. One can calculate the mean supercoiling 
density by differentiation of the free energy density 
\begin{eqnarray}
\label{eq:sigma_from_F}
\langle \sigma \rangle = -\frac{1}{\omega_0} \, 
\frac{\partial \widetilde{\cal F}}{\partial \tau}.
\end{eqnarray}
The central observables of interest in this work are the mean number of 
plectonemes $\npl$ and the plectoneme length distribution $P_{\mathrm{pl}}(l)$.
The former is given by
\begin{eqnarray}
\npl &=& \frac{\sum_n n Z(n)}{Z_\text{TOT}},
\label{eq:np}
\end{eqnarray}
while the latter is found as the relative weight of those configurations
containing at least one plectoneme of length $l$. For this, we calculate
the subensemble partition function $Z(n,l)$, for states containing $n$
plectonemes of which one has fixed length $l$.  Note that this does
not affect the enthalpic term~\eqref{eq:enthalpy} nor the positional
entropy~\eqref{eq:entropy_config}. However, for this subensemble the
length partitioning entropy is now given by
\begin{equation}
	\Lambda_n^{*}(L_p,l)=\frac{n}{(n-2)!}
	\left(\frac{L_p-(n-1)l_c-l}{\Delta l}\right)^{n-2}.
\end{equation}
This is analogous to Eq.~\eqref{eq:entropy_length} as it describes the
different ways one can partition the free length $L_p-l$ among $n-1$
plectonemes. Furthermore, we included a factor $n$ as any of the $n$
plectonemes is allowed to have length $l$. Hence the partition function
is found to be
\begin{equation}
	Z(n,l)=\sum_{L_p} e^{-\beta F_{enth}(L,L_p,n)}
		\, \Omega_n(L_p)\Lambda_n^*(L_p,l),
\end{equation}
where the sums runs over multiples of $\Delta l$ for $(n-1)l_c+l\leq
L_p\leq L$. Finally, the probability of having a configuration in which at
least one of the plectonemes has length $l$ is
\begin{equation}
	P_{\mathrm{pl}}(l)= \frac{\sum_{n=1}^{n_\text{max}}Z(n,l)}{Z_{TOT}}.
	\label{eq:LengthDistribution}
\end{equation}

The theory developed here differs from prior
approaches~\cite{mark12,eman13}, by replacing the geometric description
of the plectoneme state with a phenomenological plectoneme free energy
$\tpfe$, fitted from independent umbrella sampling simulations
\cite{skor22}, which allows us to cast the description in a
much-simplified form.

An alternative theoretical approach, based on a transfer matrix (TM)
calculation, is described in Appendix~\ref{app:TM-new}.  While the TM and
MP models both show very good agreement for several physical quantities,
both are subject to distinct limitations. In the TM formalism, it is
not straightforward to impose a smallest possible length for plectonemes
$l_c$, which leads to a systematic overestimation of the mean number of
plectonemes due to the counting of unphysically small plectonemic regions.
The advantage of the TM approach, however, is that it has a simpler
analytical structure than the MP model.  As a consequence, this model
permits the analytical description of quantities that cannot be derived
directly from the MP model.  An example these is the characteristic
plectoneme length $\xi$ (see discussion in Appendix~\ref{app:TM-new}).

\section{Monte Carlo simulations}
\label{sec:MC}

\begin{figure*}[t!]
\centering
\includegraphics[width=17.6cm]{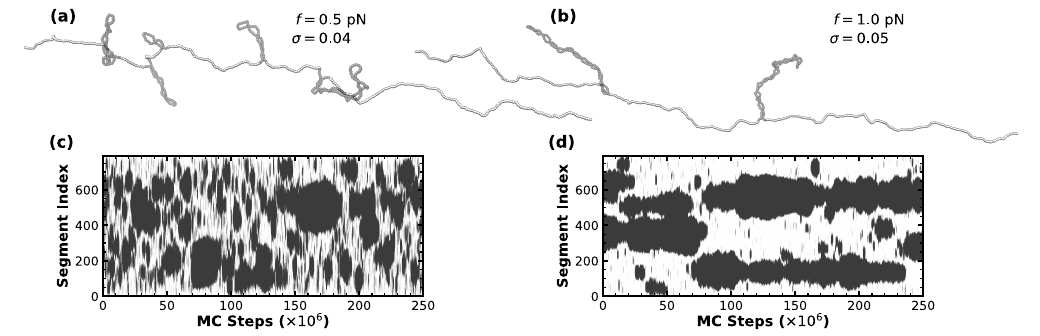}
    \caption{
    Snapshots of configurations generated with MC simulations for
    molecules of $7920$~bp (length $L=2692.8$~nm) for (a) $f=0.5$~pN,
    $\sigma = 0.04$ and (b) $f=1.0$~pN, $\sigma=0.05$.  (c) and (d):
    Kymographs showing the position of plectonemic domains (depicted as
    dark regions) for the full simulations from which the snapshots shown
    in panels (a) and (b), respectively, were taken. At $f=0.5$~pN the
    algorithm facilitates frequent changes in the location and number of
    plectonemes, while large-scale configurational changes are difficult
    to attain at $f=1.0$~pN.
    }
    \label{fig:MC-snapshot}
\end{figure*}

We performed Monte Carlo (MC) simulations based on the elastic
Hamiltonian of the twistable wormlike chain (TWLC) discretized to
$10$~bp (or $3.4$~nm) per monomer.  This model features only two free
elastic parameters: the bending modulus $A$ and the twist modulus
$C$, which—following the general choice in this work—were set
to $40$~nm and $100$~nm, respectively. Electrostatics and steric
hindrance are modeled via hard sphere potentials of radius $2$~nm
associated with each monomer.  This radius was chosen to mimic the
behavior of DNA under physiological ionic conditions~\cite{rybe93},
i.e., roughly $150$~mM monovalent salt. In previous work, this
hard-sphere radius in conjunction with the mentioned elastic constants
was found to yield the best agreement with experimental single-molecule
measurements~\cite{vand22}. Configurations are generated via a series of
cluster moves as described in~\cite{vand22}. Simulations are conducted
in the fixed linking number ensemble. This is facilitated by detecting
and subsequently rejecting topology-violating moves, and by including
repulsion planes (via hard-wall potentials) normal to the force director
field, which move in tandem with the chain termini. Further details
about the model are provided in Refs.~\cite{skor22} and \cite{vand22}.

All simulations were conducted for molecules of length $7920$~bp, which
translates into $792$ beads at the aforementioned resolution and a
contour length of $L=2692.8$~nm. Six different forces were considered,
ranging from $0.4$~pN to $1$~pN.  For each force, we examined a range
of supercoiling densities, from zero to values well beyond $\sigma_p$,
the supercoiling density at which the entire chain is expected to be
in the plectonemic phase. Configurations from two such simulations are
depicted in Fig.~\ref{fig:MC-snapshot}(a) and (b).

Accurate sampling of quantities such as the mean number of plectonemes
and the plectoneme length distribution requires generating uncorrelated
configurations.  At high supercoiling densities, where most of the
molecular length is contained within plectonemic regions, rearranging
plectonemes along the chain requires substantial confluence of individual
Monte Carlo moves.  Consequently, sampling—especially at large
forces, where plectonemic coiling is tight—becomes very inefficient.
Simulations effectively remain stuck in given configurations for
a large number of MC moves.  The decrease in sampling efficiency
due to enhanced coiling density induced by the force is illustrated
in the kymographs of Figs~\ref{fig:MC-snapshot})(c) and (d), which
display the location of plectonemes as shaded regions.  For $f=0.5$~pN
(\Figref{fig:MC-snapshot}(c)) the MC algorithm refreshes the distribution
of plectonemic regions along the chain quite frequently, while the
$f=1$~pN (\Figref{fig:MC-snapshot}(d)) simulation exhibits only a handful
of large-scale rearrangements.

To boost sampling efficiency, we employed an enhanced sampling method we
term topological replica exchange sampling. Simulations of all considered
linking numbers corresponding to the same force are run concurrently.
At regular MC step intervals, attempts are made to exchange configurations
between neighboring linking number states. The linking number in the
two corresponding configurations is adjusted uniformly by modifying the
twist to fit the respective ensemble. Exchanges are accepted based on the
usual Metropolis criterion. This type of configuration exchange allows
highly congested configurations to migrate into low supercoiling density
states, where configurational refreshing is efficient. Even with enhanced
sampling, simulations of forces beyond $1$~pN yielded poor statistics,
which prompted us to limit the range of considered forces to this value.

\subsection{Evaluation of Monte Carlo Simulations}

Plectonemic regions are identified based on their contribution to the
total writhe of the configuration. Writhe is a quantity which measures the
amount of coiling of a closed curve around itself.  Large writhe density
is therefore a differentiating property of plectonemic supercoils,
setting them apart from the comparably low writhe density stretched
phase. Mathematically, writhe is expressed as a double integral along a
closed curve \cite{full71,schi21}.  In the context of discrete chains,
this can be reduced to a double sum \cite{klen00,skor22}
\begin{eqnarray}
    \text{Wr} = \sum_{i=1}^N \sum_{j=1}^N  \omega_{ij},
\end{eqnarray}
over the pairwise contributions $\omega_{ij}$, representing double
line integrals over pairs of straight segments (see Ref.~\cite{klen00}
for details).  We refer to previous work for more detailed description of
the plectoneme detection algorithm. As a final condition, we classify only
those regions as plectonemes that contribute at least one unit of writhe.

\begin{figure*}[ht]
    \centering
    \includegraphics[width=17.6cm]{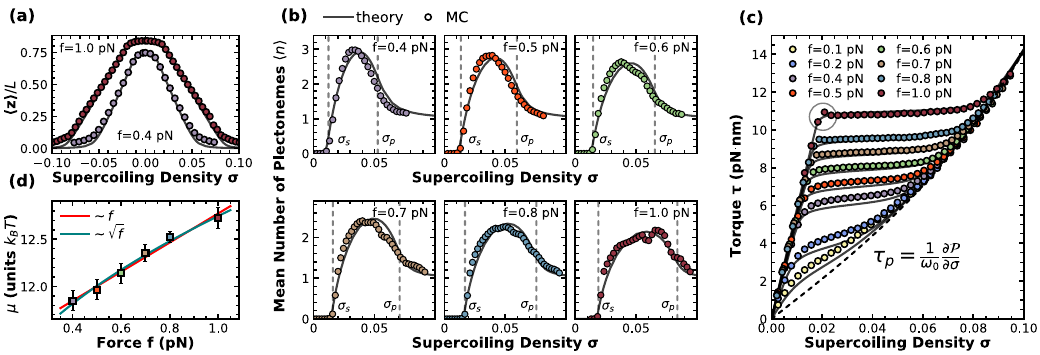}
    \caption{(a-c) Summary of comparison of MC simulation data
    (circles) with the MP model of Sec.~\ref{sec:theory} (solid lines)
    for various quantities.  Six different forces were simulated. The
    symbol colors in (a) and (b) follow the legend given in (c). (a)
    Average relative extension $\langle z \rangle /L$ vs $\sigma$.  (b)
    Average number of plectonemes $\langle n \rangle$ vs $\sigma$. (c)
    Average torque $\langle \tau \rangle$ vs $\sigma$.  The plectoneme
    nucleation free energy $\mu$ is a free parameter in the MP model which
    can be obtained from fitting Eq.~\eqref{eq:np} to data of the average
    number of plectonemes (b).  A plot of $\mu$ vs force is shown in (d).}
    \label{fig:plecto-number}
\end{figure*}

Torque can be directly measured in simulations by constraining the
rotational state of the last bead within a harmonic potential, similar to
the torsional traps used to measure torque in magnetic torque tweezer
experiments~\cite{lipf10}.  This method allows for precise control
and measurement of torque.  Alternatively, the average torque can be
calculated from the twist strain using the formula:
\begin{eqnarray}
    \label{eq:torque_measurement}
    \tau = \frac{2 \pi  \kt C}{L} \langle \Delta \mathrm{Tw} \rangle,
\end{eqnarray}
where $\Delta \mathrm{Tw}$ is the accumulative excess twist along the
molecule.  Experimentally, one is limited to the readout and manipulation
of the magnetic bead, which does not provide access to the precise
twist state of the molecule.  In contrast, our simulations yield detailed
information about the twist state and other properties, allowing us
to deduce these by direct observation.

\subsection{Number of plectonemes and Torque}

Data sampled with the MC simulations for the extension, mean number
of plectonemes, and torque, for the range of considered forces and
supercoiling densities, are displayed in Figs.~\ref{fig:plecto-number}(a),
(b), and (c), respectively. The extension exhibits the well-documented
quadratic decrease for small supercoiling densities—as described by the
theory of chiral entropic elasticity~\cite{moro98}—followed by a nearly
linear decrease in the post-buckling regime related to the conversion of
stretched phase to plectonemes.  Plectonemes are most numerous at low
forces, see Fig.~\ref{fig:plecto-number}(b).  For any given force, the
number of plectonemes varies non-monotonously with increasing supercoiling
density.  As expected, no plectonemes are observed for small $\sigma$
(below the buckling point $\sigma_s$).  Past the buckling point, this
number steadily increases until a maximum is reached at a point roughly
halfway through the post-buckling regime.  This point corresponds to
an extension reduction of about $50~\%$ relative to the relaxed state
of the corresponding force ($\sigma=0$).  Beyond this maximum, $\npl$
decreases steadily until eventually converging to $\npl \to 1$ in the pure
plectonemic phase.  As outlined before, the best sampling is achieved
for small forces and low supercoiling densities, while sampling becomes
unreliable past the point of maximum number of plectonemes for $f=1$~pN.

The torque varies linearly with $\sigma$ in the pre-buckling regime,
see Fig.~\ref{fig:plecto-number}(c). Post-buckling, the torque response
remains nearly flat at higher forces but shows a pronounced force
dependence at lower forces. To highlight this force dependence,
we included simulation data for forces as low as $0.1$~pN. MT
studies have reported an abrupt buckling transition, where the
average extension sharply drops at the buckling point, and a torque
`overshoot'~\cite{fort08,ober12}, both of which have been predicted by
both theory and simulations~\cite{mark12,brut10}.

In our MC data, we observed a tiny signature of such torque overshoot
at the highest force analyzed $f=1.0$~pN, see encirclement in
Fig.~\ref{fig:plecto-number}(c). The torque overshoot is related to
the energy barrier crossing of plectoneme nucleation and is therefore a
finite-size effect. Accordingly, it is most visible for short molecules
($< 2$~kbp) and strong forces. Lastly, we note that for all considered
forces the torque appears to converge to a single curve once approaching
the fully plectonemic state.


To compare the MP model to the MC simulations we invoke higher-order
corrections to the free energy densities of stretched and plectonemic
phase (Eqs.~\eqref{eq:defS} and \eqref{eq:defP}). These corrections
introduce two additional parameters that have been determined
by independent free energy calculations in previous work, see
\appref{app:quartic} and \cite{skor22} for a deeper discussion.

The only remaining free parameters in the model are the smallest possible
plectoneme size $l_c$ and the plectoneme nucleation free energy $\mu$,
both of which can be determined from MC data.  The value of $l_c$
can be estimated from plectoneme length distributions obtained
from MC simulations which will be discussed in detail in Section
\ref{sec:LengthDistribution}.  Once $l_c$ is fixed as a function of force,
$\mu$ can be determined by fitting the model to the MC data.

In principle, any observable can be used for the fit, however, we find the
number of plectonemes to be most sensitive to $\mu$ therefore yielding the
lowest uncertainty for the fitted parameter.  The extension is found to
be almost entirely insensitive to $\mu$, while torque is only sensitive
to $\mu$ for small forces, where the high-force expansion free energies
(Eqs.~\eqref{eq:gf_TWLC} and \eqref{eq:af_TWLC}) become unreliable.
There is, however, a caveat in employing $\npl$ as a basis for fitting
$\mu$ in that the MC results for $\npl$ inadvertently depend on the cutoff
writhe used to classify plectonemes. This introduces an additional layer
of uncertainty on $\mu$. Regardless, $\npl$ yields the most direct access
to $\mu$. As described above, we consistently used a cutoff value of one
unit of writhe to classify plectonemes.  Furthermore, numerical values
of $\mu$ always defer to a particular choice of the discretization
length $\Delta l$. Changing the discretization length from $\Delta l$
to $\Delta l'$ in turn changes $\mu$ to $\mu'$ following

\begin{equation}
    \label{eq:rescaling_mu}
    \mu'=\mu+2\kt\log\left(\frac{\Delta l}{\Delta l'}\right).
\end{equation}
In the above, the second term takes into account the gain and loss
of entropy when changing the mesh size of the system.  The resulting
values of $\mu$ for individual stretching forces with $\Delta l =
1$ nm are displayed in \Figref{fig:plecto-number}(d).  Previous work
suggests $\mu$ to scale as $\sqrt{f}$~\cite{mark12,eman13}.  However,
the relatively small range of considered forces does not permit us to
discern this behavior from linear scaling. We, therefore, included fits
for both types of scaling.

\begin{figure}[t!]
    \centering
    \includegraphics[width=8.6cm]{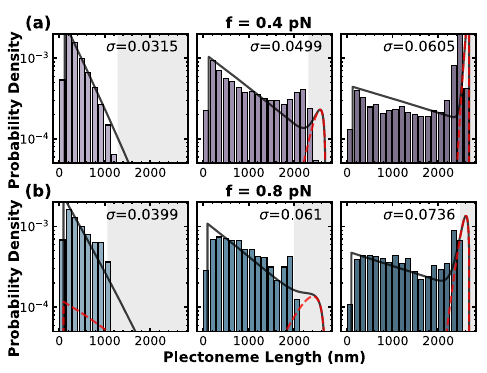}
    \caption{(a,b) Comparison of plectoneme length distributions
    from MC simulations (bars) and MP model as given
    by Eq.~\eqref{eq:LengthDistribution} (solid lines) for three
    different $\sigma$ and (a) $f=0.4$~pN, (b) $f=0.8$~pN. For small
    $\sigma$, plectonemes are short and the length distributions are
    rapidly decaying. For large $\sigma$, plectonemes are long and the
    distributions are peaked at lengths close to the total length of
    the DNA molecule. MC data hardly extend to lengths beyond $L_p$
    (grey area), given in Eq.~\eqref{def:lp}. Distributions calculated
    with the MP model extend to the entire length of the molecule.
    This difference is due to the different ensembles used in the two
    calculations, see discussion in the text.} 
    \label{fig:distributions}
\end{figure}

Observables calculated with the MP model using values of $\mu$ fitted
from the average number of plectonemes are found to incorporate
all the aspects observed in the MC data, short of the small torque
overshoot for the $1$~pN data (marked by a small grey circle in
Fig.~\ref{fig:plecto-number}(c)). As shown in the torque vs $\sigma$ graph
(Fig.~\ref{fig:plecto-number}(c)), the theory captures the deviations
from a first-order transition that manifests in the non-constant torque in
the coexistence region, corresponding to $\tau^*$ of Eq.~\eqref{eq:tau*}
in the quadratic model.  For all forces, the torque curves are found to
eventually converge to the torque curve of the plectonemic phase.  This
quantity can be obtained via differentiation of the plectoneme free energy
and is shown as a dashed line in Fig.~\ref{fig:plecto-number}(c). The
curvature of $\tau_p$ is a direct indication for the breakdown of the
quadratic approximation for ${\cal P}(\psi)$ and justifies the requirement
of a quartic free energy extension, see \Eqref{eq:plec_quart_fe}.

\subsection{Plectoneme Length Distribution}
\label{sec:LengthDistribution}

Plectoneme length distributions obtained from MC simulations (bars) and
the MP model (solid lines) are shown in Figs.~\ref{fig:distributions}(a),
and (b). The data correspond to three separate supercoiling densities,
$\sigma$, for two different forces: (a) $f = 0.4$~pN and (b) $f =
0.8$~pN.  The theoretical curves of the MP model are plotted using the
corresponding values of $\mu$ (see Fig. \ref{fig:plecto-number}(d))
while $l_c$ is obtained from the sharp drop of the plectoneme length
distributions obtained for small plectoneme lengths (see Appendix
\ref{app:MinimalLength} for details).  While the theory predicts a
non-zero probability of finding plectonemes of the longest possible
length, i.e., the length of the entire molecule, the simulations
exhibit a sharp probability dropoff around a particular and supercoiling
density-dependent length. This discrepancy stems from the difference
in the considered ensemble. The simulations are conducted in the fixed
linking number ensemble, where for a given $\sigma$ the length fraction
stored in the plectonemic state is expected to be~\cite{mark07,skor22}
\begin{equation}
    \label{def:lp}
	L_p = (1 - \langle \nu \rangle) L = 
	\frac{\sigma - \sigma_s}{\sigma_p - \sigma_s} \, L.
\end{equation}

Unless $\sigma$ is close to $\sigma_p$, the probability of $L_p$
to fluctuate to $L$ is exceedingly unlikely (for details see e.g.,
Ref.~\cite{skor22}). The theory, on the other hand, is derived in the
fixed torque ensemble which permits linking number fluctuations. Since
torque differences are generally small (except for the smallest of forces;
see \Figref{fig:MC-snapshot}(c)), linking number fluctuations can be
large, allowing plectoneme length fluctuations over the entire possible
range. The grey shaded region in Figs.~\ref{fig:distributions}(a),
and (b), indicates plectoneme lengths above $L_p$, confirming
Eq.~\eqref{def:lp} to provide an upper bound to plectoneme length
fluctuations observed in MC simulations.

For plectoneme lengths below $L_p$ we find reasonable agreement between
the MC data and the MP model. For relatively short plectonemes, the theory
predicts an exponential plectoneme length distribution. The corresponding
decay length $\xi$ constitutes the characteristic plectoneme length. While
the exponential decay cannot be directly derived from the MP model due
to its complex analytical structure, it can be obtained analytically
using the TM approach (see Appendix~\ref{app:TM-new}).  The predicted
characteristic plectoneme length from the TM theory is in line with the
results obtained from MC simulations.

For sufficiently large $\sigma$ both theory and simulation display a peak
in the probability density close to the largest observed length. This
peak can be explained by considering plectoneme length distributions
within subensembles of a fixed plectoneme number. The dashed line shows
the corresponding distribution for the subensemble containing only a
single plectoneme, which illustrates that the peaks are a result of
these subensemble states.

\section{Comparison with experiments}
\label{sec:expt}

\begin{figure}
    \centering
    \includegraphics[width=8.6cm]{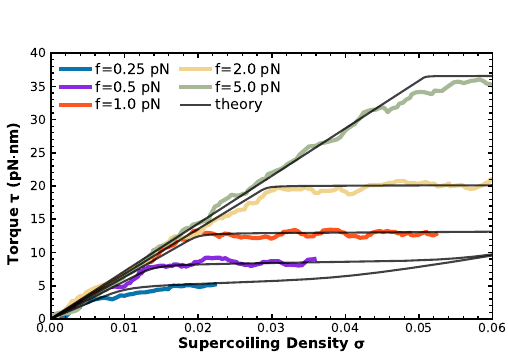}
    \caption{Comparison between experimental measurements (symbols)
    of torque as function supercoiling density and the results
    from the MP model (solid lines), as obtained by inverting
    Eq.~(\eqref{eq:sigma_from_F}). Experimental data are from
    \cite{gao21}, courtesy of M. Wang.}
    \label{fig:AOT}
\end{figure}

The MC simulations used in this work were parametrized to match
linking number-dependent surface-bead distances and variances measured
with single-molecule magnetic tweezers for $7.9$~kb DNA tethers,
see~\cite{vand22} for details. An interesting prediction of the theory
and the MC simulations is the non-flat torque response observer for low
forces in particular (also predicted by Emanuel et al.~\cite{eman13}).
We compare these results with optical tweezer measurements of DNA tethers
reported by Gao et al.~\cite{gao21}. Figure~\ref{fig:AOT} shows the
relation of torque vs. supercoiling density for forces ranging from
$0.25$~pN to $5$~pN. The model developed in this paper (black line) is
co-plotted with the experimental data (colored lines).  To produce the
solid lines of Fig.~\ref{fig:AOT} the MP model with the same parameters
as in the rest of the paper were used, except for the coefficient $P_2$
which describes the quadratic term in the plectoneme free energy,
see Appendix~\ref{app:quartic}. This was set to $P_2 = 20$~nm in
Fig.~\ref{fig:AOT}, which is higher than the value $P_2 = 14.3$~nm used
in the rest of the paper. Notably, the value of $P_2=20$~nm coincides
with the one reported by Gao et al.~\cite{gao21}, which was obtained
by fitting other torsional data.  While it is difficult to observe a
signature of multiplectonemes from the higher forces, the $f=0.25$~pN
data is found to closely align with the MP model. Unfortunately, due to
technical issues arising when the magnetic bead comes into proximity with
the flow cell surface, the experimental data is limited to comparably
small values of $\sigma$. It would be interesting to experimentally
verify the torque response in a wider range of supercoiling density to
corroborate the convergence into a single torque curve as predicted by
the theory. Such measurement would give direct access to the free energy
landscape of the plectoneme phase.

\begin{figure}
    \centering
    \includegraphics[width=8.6cm]{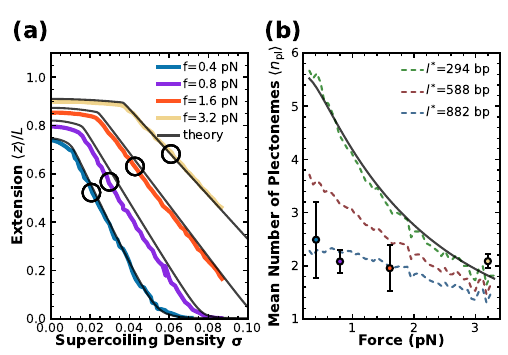}
    \caption{(a) Comparison between experimental force-extension curves
    (symbols) taken from \cite{vanl12} and results obtained from
    the MP (solid lines) for a $21$~kb DNA.  The circles indicate the
    conditions under which the fluorescence measurements were performed.
    (b) Mean number of plectonemes as function of force obtained from
    fluorescence microscopy measurements \cite{vanl12} and according
    to the MP model from Eq. ~\eqref{eq:np}. Dashed lines are obtained
    from Monte Carlo sampling the MP model (MC-MP) in conjunction with
    a minimal plectoneme detection length $l^*$ (thus plectonemes of
    lengths $l > l_c$ are generated, but only those with length $l > l^*$
    are counted in the calculation of $\npl$). Results indicate that the
    experimental resolution of \cite{vanl12} is limited to plectonemes
    of about 1~kb in length.}
    \label{fig:Loenhout}
\end{figure}

Lastly, we turn to a set of experimental data that gives direct access
to plectoneme statistics. Van Loenhout et al. \cite{vanl12} developed an
MT setup that allows for the sideway pulling of fluorescently labeled
DNA molecules and the subsequent visualization of the DNA contour via
epi-fluorescence. Plectonemes appear as bright spots as they correspond
to regions of high DNA density. Using this setup, it is possible to
locate plectoneme positions and follow their dynamics in real-time. The
quoted study used $21$~kb DNA molecules. We first verify the validity of
our theory vis-\`a-vis the experimental data by comparing measured and
theoretical extension as a function of applied force and supercoiling
density, see Fig.~\ref{fig:Loenhout}(a)). The theoretical curves were
constructed with the same parameters as used to compare to MC data. The
observed agreement attests to the quality of the model parametrization.

Average plectoneme numbers $\npl$ vs applied force are shown in
Fig.~\ref{fig:Loenhout}(b).  Experimentally considered supercoiling
densities were specifically selected such that 25\% of the
DNA length is contained in the plectonemic phase (see circles in
Fig.~\ref{fig:Loenhout}(a)). We used the MP model to compare the average
number of plectonemes $\npl$ to the experimental data of \cite{vanl12}.
Values of $\mu$ and $l_c$ are chosen by extrapolating the $\sqrt{f}$-fit
from \Figref{fig:plecto-number}(d) and \Figref{fig:minimal_size}(b)
respectively.  To ensure comparability to experiments, the MP model
was evaluated at torques for which $L_p$ amounted to 25\% of the
DNA molecule's contour length.  Fig.~\ref{fig:Loenhout}(b) indicates
the theory (solid line) to predict a considerably larger number of
plectonemes for the experimentally considered molecule length and
force and supercoiling density conditions than experimentally observed.
Moreover, the MP model predicts the mean number of plectonemes $\npl$
to decay rapidly with the applied tension $f$, while experimental
observation suggests much lower force-dependence.

This discrepancy might be attributed to the limited experimental
resolution, which does not allow for the observation of short plectonemes.
To further investigate the typical plectoneme detection limit in the
experiment, we sampled MP model configurations using a Monte Carlo
algorithm (referred as MC-MP).  This algorithm starts from a DNA chain
which is fully in the stretched phase.  Its configuration is updated
by either randomly selecting a single site and changing its state from
the S(P)-phase to P(S)-phase, or by selecting a range of sites and
changing their states collectively. Moves that generate plectonemic
domains smaller than $l_c$ are immediately rejected. Otherwise, changes
made are accepted/rejected according to the Metropolis rule. The dashed
colored lines in Fig.~\ref{fig:Loenhout}(b) are MC-MP simulation data
in which a cutoff length $l^*$ is used to detect plectonemes, meaning
that only plectonemes of length longer than $l^*$ are considered in the
determination of $\npl$.  As shown in Fig.~\ref{fig:Loenhout}(b), the
MC-MP data is in reasonable agreement with experiment when a minimal
plectoneme detection length of $l^* = 0.9$~kb is used.  Under these
conditions the theory predicts the average number of plectonemes to be
weakly dependent on force as is apparent from the experimental data. This
potentially suggests the fluorescence measurements to be sensitive only to
plectonemes of size larger than approximately $\sim 1$~kb, while shorter
plectonemes may evade detection. This effect would be most prominent
for small forces, where plectonemes are preferentially short and numerous.


\section{Conclusion}
\label{sec:conclusion}

In this paper, we presented the results of MC simulations of stretched
and torsionally constrained DNA, modeled as a twistable wormlike
chain. DNA molecules were represented as a series of coarse-grained beads
(each corresponding to 10~bp) characterized by bending and torsional
stiffnesses, chosen to reproduce experimental data of bending and
torsional persistence lengths. We analyzed the statistical properties
of plectonemes, which form along the molecule once it is over- or
underwound. Using an algorithm developed in prior work~\cite{skor22}, we
detected the number of plectonemes and their lengths for configuration
snapshots from a MC-generated equilibrium ensemble at fixed linking
number and stretching force.  In parallel, we developed a statistical
mechanical model of stretched supercoiled DNA which we referred to as MP
model.  This model is a simple extension of the two-phase model of Marko
\cite{mark07}.  An excess free energy term is introduced as the penalty
for the interface between a stretched DNA and a plectonemic domain.
All parameters of the MP model, except the phase boundary free energy
and plectoneme length threshold were determined in prior studies that
focused on other properties of supercoiled DNA \cite{vand22,skor22}. The
two-phase model was proven to capture several features of stretched and
torsionally constrained DNA \cite{mark07} including extension fluctuations
\cite{vand22,skor22}. Here we showed the model to quantitatively reproduce
multiple features observed in simulations, such as the average number of
plectonemes and their length distribution. The agreement with simulations
is remarkable as the model contains only two adjustable parameters which
makes it more accessible than more sophisticated approaches proposed in
the prior literature \cite{mark12,eman13}. The MP model is a valuable
tool to infer physical properties of stretched and supercoiled DNA,
which are either not accessible in experiments or would require long
and computationally costly simulations.


\begin{acknowledgments}
Discussions with Pauline Kolbeck, Jan Lipfert, Willem Vanderlinden
are gratefully acknowledged.  MS acknowledges financial support from
Fonds Wetenschappelijk Onderzoek-Vlaanderen (FWO) Grant No.~11O2323N.
ES acknowledges financial support from Fonds Wetenschappelijk
Onderzoek-Vlaanderen (FWO) Grant No.~1SB4219N.  ES and HS were supported
by the Deutsche Forschungsgemeinschaft (DFG, German Research Foundation)
under Germany’s Excellence Strategy -- EXC-2068 -- 390729961.
\end{acknowledgments}

\appendix

\section{Higher order extensions}
\label{app:quartic}

For the analysis of simulation data, we extended two features of the
free energies \eqref{eq:defS} and \eqref{eq:defP}.  We added a quartic
term in the supercoiling density $\psi$ of the plectonemic free energy,
which was extended as follows
\begin{equation}
    \pfe(\psi) = \left(\frac{P_2}{2}\psi^2  + 
\frac{P_4}{4} \psi^4 \right) \omn^2 \kt.
    \label{eq:plec_quart_fe}
\end{equation}
The coefficients $P_2$ and $P_4$ were fitted to MC simulation data in
\cite{skor22}, using umbrella sampling and suitable boundary conditions
to induce a pure plectonemic phase. As the supercoiling density $\psi$
is dimensionless $P_4$ and $P_2$ in \eqref{eq:plec_quart_fe} have
units of length.  Fitted free energies give $P_2=14.4\pm0.3$~nm and
$P_4 = 520\pm30$~nm~\cite{skor22}.  It was not necessary to extend the
stretched phase free energy ${\cal S}(\phi)$ given in \eqref{eq:defS},
beyond the quadratic term in $\phi$. This is because stretched segments
attain a maximum value $\phi \approx \sigma_s \ll \sigma_p$.  The need
for a quartic term in \eqref{eq:plec_quart_fe} arises from the fact that
plectonemes are characterized by a large supercoiling density.  However,
we extended Eq.~\eqref{eq:gf_TWLC} to the next order in the force $f$
to better account for the low force regime.  We used the following
expansion \cite{skor22}
\begin{equation}
    g(f) = f\left( 1 - \sqrt{\frac{\kt}{Af}} + g_2 \frac{\kt}{Af} \right).
\label{eq:g2}
\end{equation}
Ref.~\cite{skor22} reports $g_2=0.3$, obtained from the numerically
exact solution of a stretched wormlike chain \cite{mark95} where
we took $A=40$~nm and $\kt=4.1$~pN$\cdot$nm, corresponding to room
temperature. Note that the term proportional to $g_2$ contributes to the
stretched phase free energy as an overall force-independent constant. This
is important for the double tangent construction as it gives a constant
relative shift to the stretched and plectoneme phase free energies.

\section{Transfer Matrix approach}
\label{app:TM-new}

An alternative approach to the multiplectonemic phase free energy is
based on the following $2 \times 2$ transfer matrix (TM)
\begin{equation}
\label{def:TM}
    T =
    \begin{pmatrix}
      s & wp \\
      ws& p
    \end{pmatrix},
\end{equation}
where $s$ and $p$ are the weights of the two phases
\begin{eqnarray}
s &=& \exp(-\beta \, \Delta l \, \tsfe),
\n
p &=& \exp(-\beta \, \Delta l \, \tpfe),
\label{def:sp}
\end{eqnarray}
corresponding to a discretization length $\Delta l$. The term $w < 1$
is the additional weight associated with the interface between stretched
and plectonemic segments. This is linked to the interfacial free energy
$\mu$ of \eqref{eq:enthalpy} through the relation
\begin{eqnarray}
  w &=& e^{-\beta \mu/2},
\label{app:w}
\end{eqnarray}
assuming the same choice of $\Delta l$ as in the multiplectoneme model.
For different choices of $\Delta l$, $\mu$ rescales according to
Eq.~\eqref{eq:rescaling_mu}, see discussion below.

This TM approach is identical to the Zimm-Bragg model used to describe
helix-coil transition in polypeptides \cite{dill10} and is also often
used to describe two-state transitions of single molecules; see e.g.,
\cite{viad21} for a recent example. The TM matrix multiplication
generates the Boltzmann weights of all possible configurational
permutations of stretched and plectonemic segments. For example, a
particular configuration containing two plectonemes of size $m\Delta l$
and $q\Delta l$ has weight
\begin{equation}
    \label{eq:boltz}
    s^k w p^m w s^{r} w p^q w s^t.
\end{equation}
We impose two stretched segments at the two ends of the molecule. This
choice of boundary condition reflects the MT setup in which the tethering
of the DNA to bead/surface imposes certain constraints inhibiting
the nucleation of plectonemes at the DNA termini. The corresponding
stretched-stretched partition function for a molecule of length $N \Delta
l$ is given by
\begin{eqnarray}
\label{app:Zss}
    Z_{ss} (N) &=& 
	s
    \begin{pmatrix}
        1 & 0 
    \end{pmatrix}
    T^{N-1} 
    \begin{pmatrix}
        1 \\
        0
    \end{pmatrix}
\nonumber\\
     &=&  \frac{s- \lambda_-}{\lambda_+ - \lambda_-}\lambda_+^N 
    + \frac{\lambda_+ - s}{\lambda_+ - \lambda_-}\lambda_-^N,
\end{eqnarray}
where $\lambda_{\pm}$ are the eigenvalues of $T$
\begin{eqnarray}
\label{eq:def_lambda}   
\lambda_\pm &=& \frac{1}{2} \left\{
s+p \pm \sqrt{(s-p)^2+4 w^2 s p} \right\}.
\end{eqnarray}
To obtain \eqref{app:Zss} we 
used the decomposition
\begin{eqnarray}
    \begin{pmatrix}
        1 \\
        0
    \end{pmatrix} 
    &=& \frac{1}{wp} \left[
    \frac{s-\lambda_-}{\lambda_+ - \lambda_-} \vec{v}_+ + 
    \frac{\lambda_+ - s}{\lambda_+ - \lambda_-} \vec{v}_-
    \right],
\end{eqnarray}
where
\begin{eqnarray}
\label{app:def_vpm}
    \vec{v}_\pm &=&
    \begin{pmatrix}
        wp \\
        \lambda_{\pm} -s
    \end{pmatrix},
\end{eqnarray}
are the eigenvectors of $T$ corresponding to the eigenvalues
$\lambda_{\pm}$. In the TM formalism the free energy per unit of length
in the thermodynamic limit is given by
\begin{eqnarray}
\widetilde{\cal{F}} &=& \lim_{N \to \infty} \frac{-k_BT \log Z_{ss}}{N \Delta l}
=  -\frac{k_BT}{\Delta l} \log \lambda_+.
\end{eqnarray}
One can get the supercoiling density $\sigma$ by means of a torque
derivative, see Eq.~\eqref{eq:sigma_from_F}. An expression for the
average number of plectonemes can be obtained by differentiation with
respect to $w^2$, i.e.,
\begin{eqnarray}
    \npl = w^2 \frac{\partial \log Z_{ss}}{\partial w^2}.
\label{app:np}
\end{eqnarray}
The probability of finding a plectoneme of length $m \Delta l$ embedded
within a segment of stretched phase is given by
\begin{equation}
    \label{eq:length_plecto}
    P_{pl}(n) = \frac{1}{Z_{ss}(N)} 
	\sum_{k=1}^{N-m-1} Z_{ss}(k) w^2 p^m Z_{ss}(N-k-m),
\end{equation}
where $k \Delta l$ is the entry point of the plectoneme along the
curvilinear length and $Z_{ss}(k)$ is given by \eqref{app:Zss}. If
plectonemes are shorter than the whole molecule length, the distribution
decays rapidly, and the sum is dominated by the terms $N-k \gg m$ and
$k \gg m$. Approximating $Z_{ss}(k) \sim \lambda_+^k$, $Z_{ss}(N-k-m)
\sim \lambda_+^{N-k-m}$ and $Z_{ss}(N) \sim \lambda_+^N$ one finds
plectoneme sizes to be exponentially distributed
\begin{eqnarray}
    \label{eq:pleclen_exp}
    P_{pl}(m) \sim \left( \frac{p}{\lambda_+} \right)^m = e^{-m\Delta l/\xi},
\end{eqnarray}
where the characteristic length $\xi$ is given by
\begin{eqnarray}
    \label{def:xi}
    \xi = \frac{\Delta l}{\log (\lambda_+/p)}.
\end{eqnarray}
At the buckling point $\tau=\tau^*$ one has $s=p$ and therefore
$\lambda_+ = p(1+w)$. The decay length \eqref{def:xi} becomes
\begin{eqnarray}
\xi = \frac{\Delta l}{\log (1+w)} \approx \frac{\Delta l}{w},
    \label{eq:xi_at_tau*}
\end{eqnarray}
where we used $w \ll 1$, i.e., the interface between straight and
plectonemic phase has a high energy cost and thus very low probability.
The previous equation shows how the interfacial weight $w$ can be obtained
from the decay of the plectonemes length distribution at the buckling
point $\tau= \tau^*$.

To understand the discretization length dependence in Eq.~\eqref{app:w},
we can use the TM multiplication as follows. Let us consider starting from
a given discretization length $a$ and indicate with $s$, $p$ and $w(a)$
the weights corresponding to this choice of length.  The TM multiplication
$k$ times gives
\begin{eqnarray}
    T^k &\approx & \begin{pmatrix}
        s^k & {w_1}(ka)\, p^k \\
        {w_2}(ka)\, s^k & p^k 
    \end{pmatrix},
\label{eq:Tk}
\end{eqnarray}
where we kept the lowest terms in $w(a)$ giving
\begin{eqnarray}
w_1(ka) &=& w(a) \left( 1 + \frac{s}{p} + \ldots \frac{s^{k-1}}{p^{k-1}} \right),
\label{app:w1}
\\
w_2(ka) &=& w(a) \left( 1 + \frac{p}{s} + \ldots \frac{p^{k-1}}{s^{k-1}} \right).
\label{app:w2}
\end{eqnarray}
Via block decimation, the production in \eqref{eq:Tk} defines a new model
with discretization length $ka$. In this model, the interfacial weight
for going from the stretched phase to the plectonemic phase, $w_1$,
is different from the one going from the plectonemic to the stretched
phase, $w_2$.  Since all configurations compliant with the stretched phase
boundary conditions require the interfacial weights $w_1$ and $w_2$ to
appear in pairs, what matters is the average Boltzmann weight defined as
\begin{eqnarray}
w(ka) \equiv \sqrt{w_1(ka) w_2(ka)}. 
\label{app:wka}
\end{eqnarray}
Close to buckling, $\tau \approx \tau^*$, the free energies of the
stretched and plectonemic phases are degenerate, hence $s \approx p$
and $w_1 (ka) \approx w_2(ka) \approx k w(a)$ and thus
\begin{equation}
w (ka) \approx k w(a).
\end{equation}
This relation is equivalent to Eq.~\eqref{eq:rescaling_mu} and has a
simple interpretation: The multiplicative factor $k$ accounts for the
number of different positions in which the interface between stretched
and plectonemic phases can be placed in a discretized segment of length
$\Delta l = ka$. This implies that the weights of individual segments
must scale with the discretization length, $w(\Delta l) \propto \Delta l$.
Note that this dependence cancels the $\Delta l$ factor in the numerator
in \eqref{eq:xi_at_tau*}.

\begin{figure}[t!]
\centering
\includegraphics[width=8.6cm]{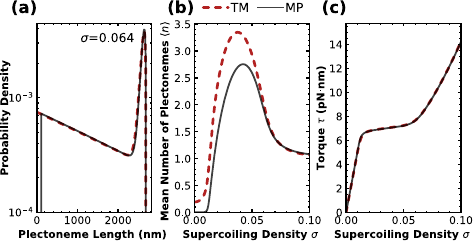}
   \caption{Comparison of the multiplectoneme model (MP, solid lines)
   and the transfer matrix model (TM, dashed line). All curves were
   calculated for the same values of the stretching force $f=0.5$~pN.
   (a) Plectoneme length distribution $P_\text{pl}(l)$ vs. length
   $l$ in lin-log scale. (b) Average number of plectonemes $\npl$ vs
   $\sigma$. (c) Torque $\tau$ vs $\sigma$.}
\label{fig:TM-cont}
\end{figure}

Figure~\ref{fig:TM-cont}(a,b,c) shows a comparison of plectoneme length
distribution $P_\text{pl}(l)$, the average number of plectonemes $\npl$
and torque vs. supercoiling density as obtained from the multiplectoneme
model (MP) and the TM formalism approach. As discussed in the text,
the former method allows a direct incorporation of a minimal threshold
length for a plectoneme $l_c$ allowing for the discrimination of short
unphysical plectonemic domains. However, in the TM formalism plectonemes
of all lengths above the discretization length $\Delta l$ are generated.
This can be seen in Fig.~\ref{fig:TM-cont}(a): the two models predict
identical length distributions, except for deviations at small $l$
where the TM approach (dashed line) predicts non-vanishing probability
down to the smallest possible length scale ($\Delta l$).  Conversely,
the introduction of the cutoff length $l_c$ excludes such small lengths
in the probability distributions calculated with the MP model (solid
line).  The TM model predicts a larger average number of plectonemes
as compared to the MP model (Fig.~\ref{fig:TM-cont}(b)), as the former
counts very short plectonemes, which are not generated in the MP model.
Both models give very close estimates of supercoiling densities $\sigma$,
see Fig.~\ref{fig:TM-cont}(c), as short plectonemes have negligible
influence on global thermodynamic quantities.

\begin{figure}[t!]
\centering
\includegraphics[width=8.6cm]{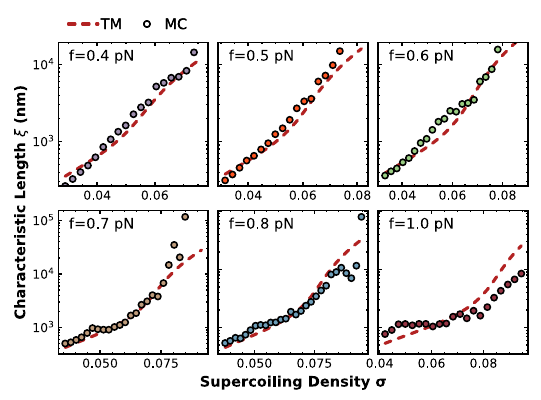}
\caption{Plots of force-dependent characteristic decay lengths $\xi$.
    Scatters indicate the values obtained by fitting the exponential
    decay of the length distributions obtained with the MC simulations
    and the dashed lines show the theoretical predictions according to
    Eq.~\ref{eq:pleclen_exp}.}
\label{fig:xi_TMvsMC}
\end{figure}

While the MP model is more appropriate to describe the physics of the
multiplectoneme phase, it has a more complex analytical structure as
it requires the calculation first of the partition function of $n$
plectonemes $Z(n)$ from \eqref{def:Zn-int} which is then summed over
all $n$ to get $Z_\text{TOT}$, see Eq.~\eqref{def:Ztot}. The TM on the
other side is simpler to handle as the thermodynamic quantities are
obtained from suitable derivatives of its eigenvalues $\lambda_\pm$.
As the TM and the MP model predict the same decay of plectoneme length
distribution we employed the latter (Eq.~\eqref{def:xi}) to compute the
decay length $\xi$ for various forces and supercoiling densities $\sigma$
and compared them with those calculated with the Monte Carlo simulations.
The results are plotted in Fig. \ref{fig:xi_TMvsMC} and show good overlap
between theory and MC simulations.

\section{Minimal plectoneme size $l_c$}
\label{app:MinimalLength}

\begin{figure}[t!]
    \centering
    \includegraphics[width = 8.6cm]{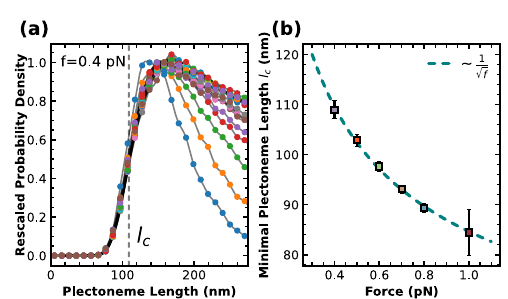}
    \caption{(a) Plectoneme length distribution for a stretching force of
    $0.4$~pN for supercoiling densities in the interval [0.021, 0.0815]
    as obtained from MC simulations (circles) along with interpolating
    cubic splines (dashed lines).  The length distributions are rescaled
    to align at their local maxima.  The minimal plectoneme length $l_c$
    is chosen as the mean-half maximum of the length distributions
    for a given force. (b) Force dependence of $l_c$.  The error bars
    indicate the standard deviation of $l_c$ determined for different
    supercoiling densities.  The dashed line indicates the result of
    fitting \Eqref{eq:emp_lc} to the deduced values of $l_c$.}
    \label{fig:minimal_size}
\end{figure}

The MP model has two free parameters, the plectoneme nucleation free
energy $\mu$ and the minimal length of a plectoneme $l_c$. Introduction of
the latter is important to avoid consideration of very short plectonemic
domains that would be unphysical as the classification of such a domain
requires the formation of at least one loop.  We determined values
of $l_c$ in a force-dependent manner based on the MC simulations.
At short lengths (see Fig. \ref{fig:minimal_size}(a)), the length
distributions rise sharply at well-determined values, independent
of supercoiling density, but depending on the stretching force. We
extract the value of $l_c$ by fitting a cubic spline through the
initial rise of the probability density and extracting the value of
the length at which $P_{\text{pl}}(l)$ reaches half of its first local
maximum. Following this procedure the force-dependence of $l_c$ was
obtained (see Fig. \ref{fig:minimal_size}(b)). In order to allow for the
inter- and extrapolation of $l_c$ we fitted the available values of $l_c$
as function of $f$ with an empirical equation of the form:
\begin{equation}
    \label{eq:emp_lc}
	l_c(f) = x_0 + x_1\sqrt{\frac{\kt}{Af}},
\end{equation}
where the numerical constants $x_0$ and $x_1$ were found to be 41~nm
and 134~nm, respectively.  We note that the scaling of $l_c$ with
$\frac{1}{\sqrt{f}}$ is in agreement with the scaling of the plectoneme
loop size \cite{mark12}.

%

\end{document}